\documentclass[a4paper,10pt]{article}
\usepackage{graphicx}
\usepackage{amsmath}
\usepackage{amssymb}
 
\begin{document}

\begin{center}
{\Large Behavior of grains in contact with the wall of a silo during the initial instants of a discharge-driven collapse}\\

\vspace{.5cm}

C. Colonnello $^a,$ 
\footnote{Corresponding author. Tel.:+58 212 9063541. Email adress: colonnello@usb.ve},
L.I. Reyes $^a$,
E. Cl\'ement $^b$
G. Guti\'errez $^a$\\

{\itshape $^a$ Departamento de F\'isica, Universidad Sim\'on Bol\'ivar, Apartado Postal 89000, Caracas 1080-A, Venezuela.}\\
{\itshape $^b$ PMMH, ESPCI, CNRS (UMR 7636) and Univ. Paris 6 \& Paris 7, 75005 Paris, France.}
\end{center}

\vspace{.5cm}

\begin{abstract}
We study experimentally gravity-driven granular discharges of laboratory scale silos, during the initial instants of the discharge. We
investigate deformable wall silos around their critical collapse height, as well as rigid wall silos. 
We propose a criterion to determine a maximum time for the onset of the collapse and find that the onset of collapse
always occurs before the grains adjacent to the wall are sliding down. We conclude that the evolution of the static friction
toward a state of maximum mobilization plays a crucial role in the collapse of the silo.
\end{abstract}

\section{Introduction}

Granular materials are essential in a vast amount of industrial processes.
Cylindrical silos are often used to store large quantities of
grains. These structures are susceptible to failure for various reasons. In particular we are interested in the case 
when the silo collapses during a gravity-driven discharge through a centered circular hole at the bottom.
It is known \cite{nedderman,gutierrez-ULA,gutierrez09,ramon-tesis} that for silos with a sufficiently thin and elastic wall there exists a 
critical height, $L_c$, of the column of grains contained in the silo such that when the initial height of the column exceeds $L_c$ the silo 
collapses during the discharge, damaging the structure permanently. When the initial height of the column is below this critical height,
deformations of the wall may be produced but they disappear by the end of the discharge. 
The existence of a critical height can be thought of in terms of classical elasticity theory of thin shells subjected to an axial stress 
\cite{gutierrez-ULA,gutierrez09,ramon-tesis,timoshenko},
where the axial stress is provided by the friction force that the grains exert on the silo.
The collapse of a silo may cause great economic losses as well as being potentially dangerous for personnel in industrial areas. It is therefore important 
to understand and prevent these catastrophic events.
The value of $L_c$ depends, among other parameters, on the diameter of the silo, 
the thickness and elastic properties of its wall, the bulk density of the granular column and the effective friction coefficient between the grains and the wall 
\cite{gutierrez-ULA,gutierrez09,ramon-tesis,gutierrez-tbp}. 
However, a complete understanding of the dynamics that leads to the silo collapse has 
not been achieved.\\
During a gravity-driven discharge, complex flow patterns arise (e.g. \cite{nguyen80,choi05}).  These flow patterns may show
stagnant zones and zones with a plug and funnel types of flow, depending on variables such as 
grain size, silo geometry and grain-to-wall friction coefficient. 
We are interested in the interaction of the grains with the wall during the initial moments of the discharge, which is when the deformation of the silo wall occurs. 
As the discharge begins only grains close to the orifice start moving, the flowing region grows until a stationary flow state is reached \cite{samadani99,arevalo07}.
During this transient state one would expect mobilization of friction, $m$, to occur locally in the contact network of the granular material, since grains
with static contacts must eventually slide as the grains enter the flowing region. Here  $m\,=\,F_f/\mu{}F_n$, with $\mu$ the effective grain-to-wall
static friction coefficient and $F_n$ and $F_f$ the magnitude of the normal and friction force between the grains and the wall.
Thus, $m$ measures how close a contact is to sliding.\\
The grains inside the silo exert an axial force on the wall through friction. This redistribution of 
the weight of a static granular column to the silo wall produces a screening effect,
that allows the pressure in the bottom of the silo to be independent on the height of the granular column when this height exceeds a certain value, called the 
screening length. This effect was first described by Janssen \cite{janssen} in 1895 and, more recently, has been the subject of several studies 
\cite{vanel99,vanel00,wambaugh10,perge12,ovarlez03,ovarlez05}.
In particular, Perge et al \cite{perge12} studied how the vertical force exerted by static grains on the silo wall increases as the grain-to-wall friction is 
mobilized. They also observed a dynamic screening effect that depends on the initial mobilization of the grains.\\
In the present article we study the behavior of the grains adjacent to the wall in cylindrical silos filled with glass beads during the initial instants of the 
discharge, before catastrophic collapse occurs. 
We study systems below and above the 
collapse threshold, observing the deformations that appear on the silo wall and the behavior of the grains adjacent to it. We want to 
determine under what conditions of friction the deformation and collapse of the silo wall are triggered. 

\section{Experimental Setup}

\begin{figure}
\centering
\includegraphics[clip, width=\textwidth]{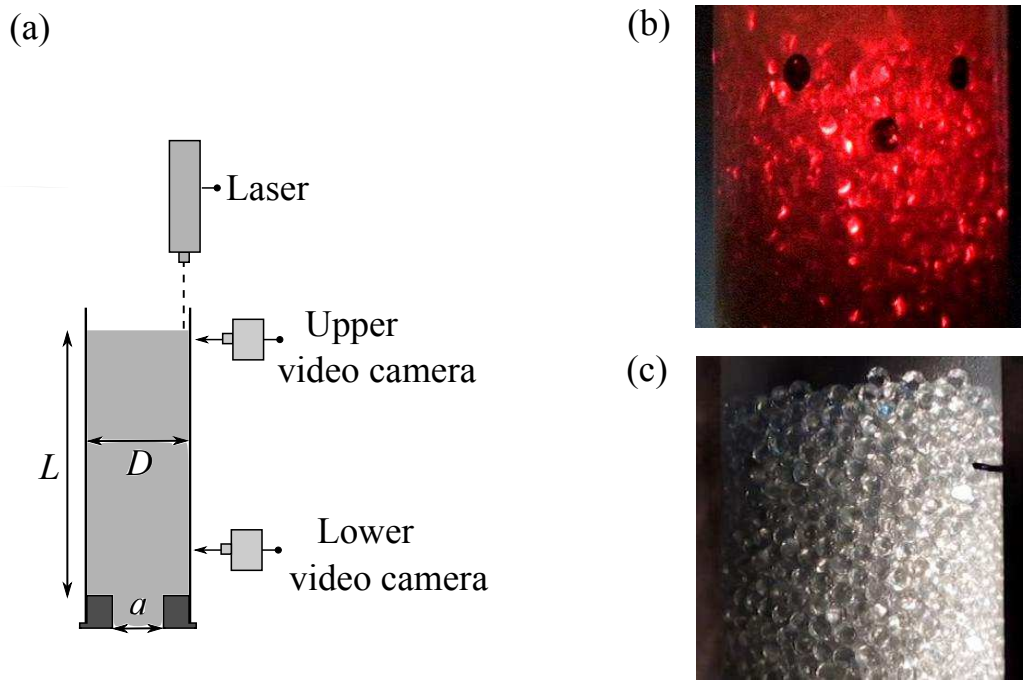}
\caption{Experimental setup. 
(a) Schematic diagram of the experimental setup. Two digital video cameras are used to monitor the movement of the grains and any deformation
around the silo wall. A laser illuminates the upper layers of grains so that one can see individual grains through translucent paper silo wall. Two mirrors (not 
depicted here) behind the silo allow us to see the deformations over all the surface of the silo. 
(b) Typical image of glass beads with diameter $d=(3.00 \pm 0.005)mm$, as seen through the translucent paper silo wall when illuminated by the laser. 
The three dark circles are reference marks on the silo wall to identify the relative movement between the grains and the wall.
(c) Typical image of glass beads with diameter $d=(3.00 \pm 0.005)mm$, as seen through a thin plastic sheet silo wall.
}\label{Exp-Setup}
\end{figure}

We work with laboratory scale cylindrical silos made of paper
and plastic sheets as well as acrylic glass (PMMA) tubes (Fig.\ref{Exp-Setup}(a)). The paper and plastic sheet silos
are made in the laboratory from a sheet wrapped around a solid cylinder and glued on a vertical narrow band, $4.0mm$ wide
\cite{gutierrez-ULA,gutierrez09}. 
The resulting tube has a diameter $D=(4.0 \pm 0.1)cm$. We used silk paper with a thickness $t=(27 \pm 1)\mu{}m$. 
This paper is translucent, making it difficult to observe the grains inside the silo (Fig.\ref{Exp-Setup}(b)), but its collapse threshold
$L_c$, as defined above, 
is easily reached in a laboratory scale experiment. 
The plastic sheet has a thickness $t=(54 \pm 1)\mu{}m$ and is totally transparent (Fig.\ref{Exp-Setup}(c)). Silos made with this material, although flexible, 
have a very high $L_c$ such that we are not able to produce their discharge driven collapse.
The PMMA silos
are also transparent and
have an internal diameter $D=(4.0 \pm 0.1)cm$ and a thickness $t=(1.0 \pm 0.1)cm$, which makes them completely rigid. 
All silos have their lower end tightly fitted around a solid cylinder, $(3.0 \pm 0.1)cm$ high, with a center hole of diameter $a$,
which is closed with a foam plug that is pulled out to produce the discharge. The  upper end
is free in all cases. The silos are filled using a centered funnel fixed at the top. The granular material 
consists of glass spheres
of diameter $d$ between $1mm$ and $3mm$.\\
We are interested in monitoring the deformation of the silo wall and the movement of the grains during the beginning of 
the discharge. To this end, we record the discharge of the silo with two digital video cameras that work at a rate of 30 fps and with a resolution of $480\times720$ pixels. 
For the transparent wall silos we use both cameras to record the position of grains near the wall in the upper and lower portions of the granular column. 
In the case of the paper silos, 
it is difficult to observe the grains through the wall. To be able to see them,
we have installed a laser which illuminates the upper layers of the granular bed 
from above, allowing us to record them on video with the upper video camera. 
The grains in the lower layers cannot be observed in paper silos with this setup as the light quickly disperses in the granular medium. 
On the other hand, the paper silos suffer deformations during grain discharge, 
which we are interested in monitoring.
These deformations occur in the lower portion of the wall, hence, for paper silos we use the lower video camera to
record the deformation process.
We have placed two mirrors forming $45^o$ behind the silo so that we can observe the whole circumference, since the deformations can form anywhere on the 
surface.
In order to determine when the discharge begins we use the soundtrack of the videos. There, a 
characteristic waveform indicates that the grains are hitting the container that collects them under the silo. 
Taking into account the time the grains spend 
in free fall, one can determine on which frame the discharge began. Finally, to monitor the movement of individual grains we use a software, 
developed using Python package SciPy, that identifies the position of individual grains through each frame of the video
using a centroid technique.
The image processing is very similar to the one used by Choi et al \cite{choi05} and Wambaugh et al \cite{wambaugh07}. 
We use the observed fluctuations in the position of the grains before the discharge to estimate the uncertainty in their position after the discharge starts
($\pm0.03d$). These are of the same order as the ones observed in the first frames after the beginning of the discharge 
(see Fig.\ref{profile},\ref{granos}).

\section{Experimental Results}

\subsection{Characterization of the onset of deformations}

\begin{figure}
\centering
\includegraphics[clip,width=\textwidth]{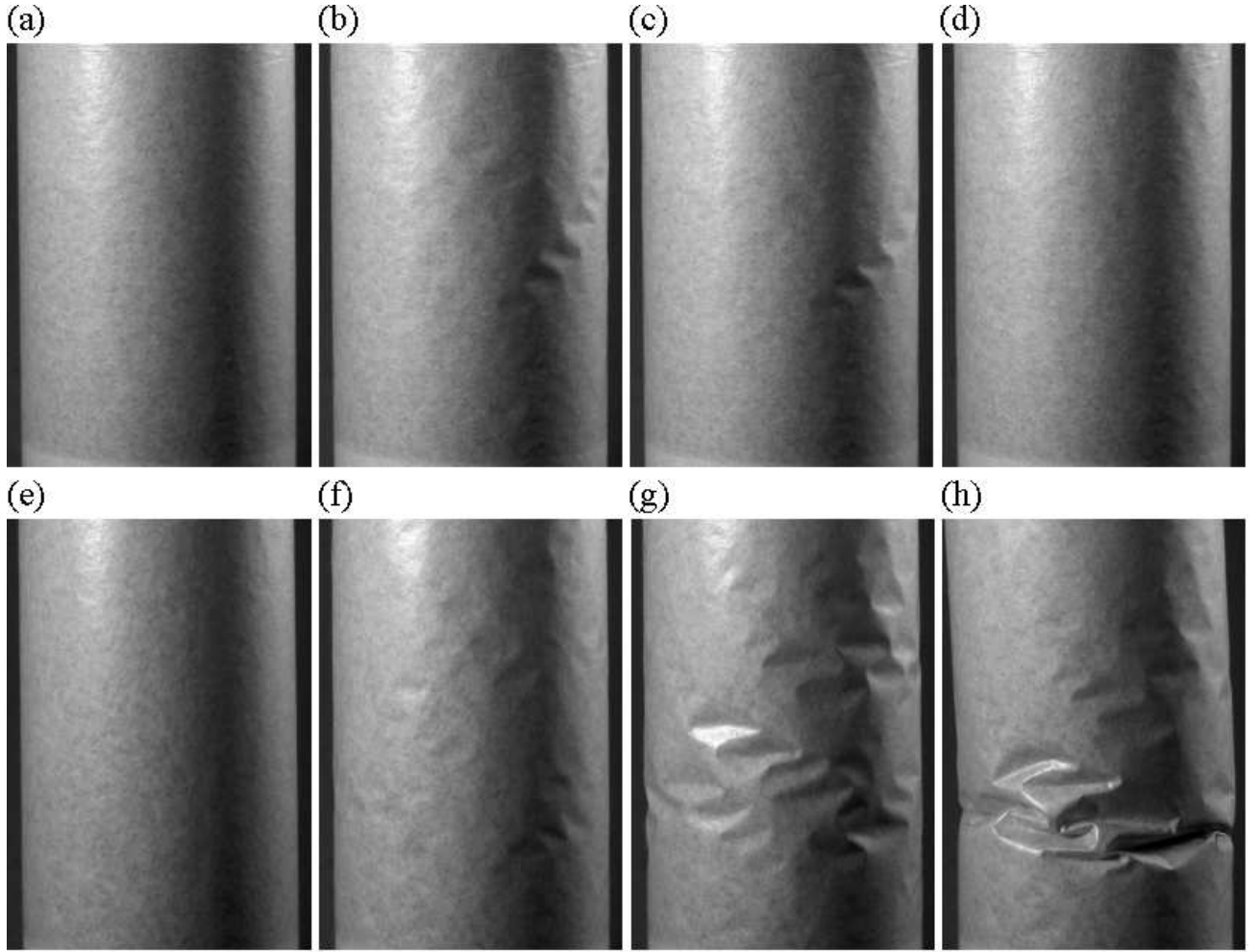}
\caption{Deformation sequence. In (a) through (d) one can observe the buckling process during a non-collapse event:
(a) silo wall before discharge, (b) localizations develop and chains or clusters of localizations are formed, (c) localizations 
start diminishing in size and disappearing as the grains flow out of the silo, (d) the silo wall returns to its initial state.
In (e) through (h) the time sequence shows a collapse event: (e) silo wall before discharge, (f) deformations develop and grow forming clusters,
indistinguishable from the ones observed in non-collapse events, (g) localizations continue growing and they start merging, 
(h) the catastrophic collapse of the silo occurs as some localizations continue to grow and become permanent
(these deformations are the ones whose history we follow back in time and whose appearance we associate with the time of onset of collapse, $\tau_c$). 
Once this state is reached the silo wall stays
permanently deformed, even after the silo is completely empty.
Permanent deformations leave wrinkles on the wall that can be clearly identified by visual inspection.
}\label{deformacion}
\end{figure}

When the cylindrical shell is subjected to axial forces, and in particular during grain discharge, diamond shaped localizations arise on the wall of the silo 
(Fig.\ref{deformacion}). These localizations often appear as isolated dimples that grow in size and multiply forming chains and clusters
around the entire circumference of the silo.
During grain discharge, each localization grows individually up to an average size, and some localizations continue growing by merging with their neighbors.
This occurs between the bottom end of the silo and a height of $2D$, for the diameter used in our experiments.
When the height of the granular
column is below $L_c$, localizations reach a stable
size and start flattening out as the grains flow out of the silo 
(Fig.\ref{deformacion}(a-d)). 
When the height of the granular column is above $L_c$, the size of some of these localizations continues to grow until the deformations become 
irreversible (Fig.\ref{deformacion}(e-h)). We consider the silo has collapsed if any deformation
can be visually identified as a permanent deformation of the wall (Fig.\ref{deformacion}(h)), that will not disappear even after all the grains 
flow out of the silo. 
We will define the time of collapse, $\tau_p$, as the time when the first permanent deformation appears on the wall. 
Usually the occurrence of one plastic deformation is followed by many such events, and the collapse of the silo results in a catastrophic failure of the 
structure. \\
We are interested in studying the conditions under which the collapse is triggered, therefore, in  
a time scale associated to the {\itshape onset} of the collapse process. 
We find that deformation processes that do not lead to the collapse and those that lead to the collapse, are indistinguishable up to the point
where some deformation becomes permanent.
Therefore, if we want to identify a time scale associated to the beginning of the collapse, the
time when the initial buckling occurs seems inappropriate. Buckling occurs in collapse and non-collapse processes and the localizations thus formed
may become stable, possibly stabilized by the internal pressure of the grains on the wall. 
In the context of classical elasticity theory \cite{timoshenko}, it is natural to expect that 
beyond a critical value of the axial stress 
we will find local deformations that will grow without bound. The appearance of these unstable local deformations signals the onset of the collapse.
Thus, we define a criterion which identifies the first localization that becomes permanent in the above sense.
Once this deformation has been identified, we look into previous frames 
of the video until we find the first frame in which it becomes visible. The time when this localization becomes visible, $\tau_c$, is an upper bound 
for the {\itshape time of onset} of the collapse. It is important to stress the fact that the time scale $\tau_c$ is an overestimate of the 
time scale associated to the onset of buckling of the wall leading to its collapse. The onset of the collapse may have been triggered before this 
localization appeared, specially since the deformation process is non-local. 

\subsection{Movement of the grains}

\begin{figure}
\begin{center}
\includegraphics[clip,width=\textwidth]{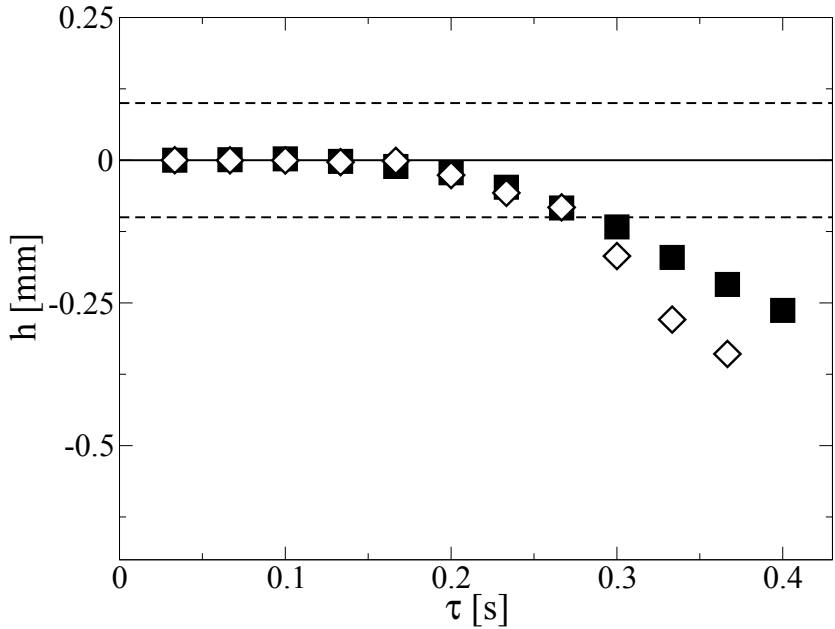}
\caption
{Time-dependent position of the grains in contact with the wall at two heights in the granular column. 
Vertical displacement of grains, $h$, as a function of time, $\tau$, at two heights in a
rigid PMMA silo filled with monodisperse glass spheres of size $d\,=\,(1.50\pm0.05)mm$ 
during a short interval of time after the discharge begins: 
grains on the upper layers of the column ($\blacksquare$) and grains around a height of $1.5D\,=\,6.0cm$ ($\Diamond$).
Silo diameter is $D\,=\,(4.0\pm0.1)cm$ and initial column height $L\,=\,(21.0\pm0.5)cm$. Within our time resolution, displacement of grains adjacent to the 
wall starts simultaneously except in a transition zone near the base of the silo. 
The solid and dashed lines represent the initial position ($h\,=\,0$) and the size of a pixel in our images, respectively ($d\sim{}15$pixels). 
The beginning of the discharge occurs at time  $\tau\,=\,0s$.}
\label{profile}
\end{center}
\end{figure}

\begin{figure}
\centering
\includegraphics[clip,width=\textwidth]{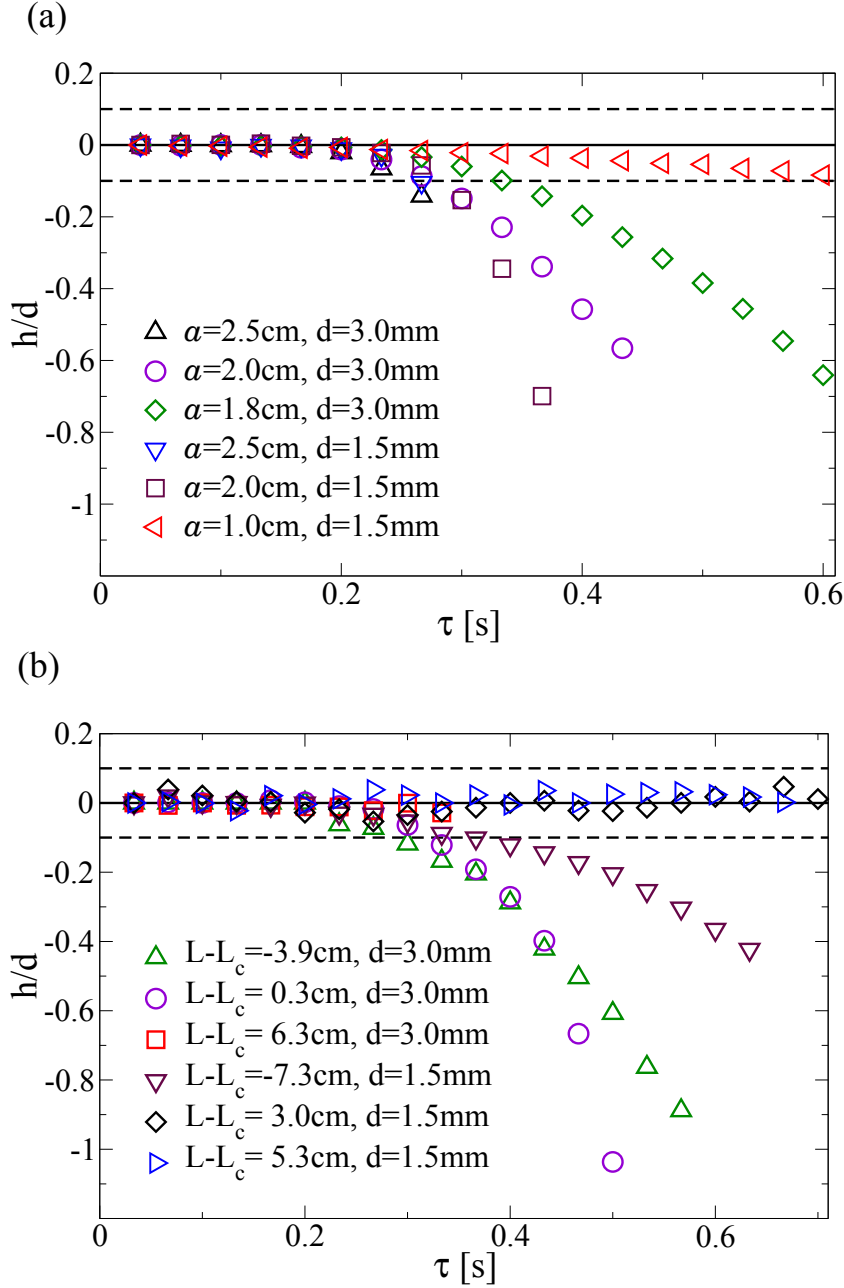}
\caption{Time-dependent position of the grains in contact with the wall of rigid and flexible silos. 
We show the average vertical displacement of the grains, $h$, as a function of time,
$\tau$, of a group of at least 10 grains in the upper layers of a granular column during the beginning of the discharge. 
(a) Rigid PMMA silos with variable discharge orifice diameter, $a$, filled with monodisperse glass spheres of two sizes:  
$d\,=\,(3.00\pm0.05)mm$ and $d\,=\,(1.50\pm0.05)mm$.
The initial filling height is $L\,=\,(21.0\pm0.5)cm$.
Notice the limited number of points for the smallest values of $d/a$. For these cases, soon after $\tau_s$ grains move too fast to be detected with the 
limited time resolution of our experiment.
(b) Paper silos filled with monodisperse glass spheres with different values of the initial filling height $L$ around the critical collapse height $L_c$.
Results are shown for two cases: particles of size $d\,=\,(3.00\pm0.05)mm$ with a discharge orifice of diameter $a\,=\,(1.8\pm0.1)cm)$, 
and particles of size $d\,=\,(1.50\pm0.05)mm$ with $a\,=\,(1.0\pm0.1)cm$. 
The critical collapse height $L_c$ is \cite{gutierrez-tbp}: $L_c\,=\,(16\pm1)cm$ and  $L_c\,=\,(19\pm1)cm$, respectively.
When $L<L_c$ no collapse occurs, but an elastic deformation of the wall may happen if $L$ is close enough to the threshold. 
When $L\geq{}L_c$ a collapse occurs and grain position is monitored only up to $\tau_p$. 
The diameter of all the silos is $D\,=\,(4.0\pm0.1)cm$. 
The solid and dashed lines represent the initial position ($h\,=\,0$) and the size of a pixel in our images, respectively ($d\sim{}15$pixels). 
The beginning of the discharge occurs at time  $\tau\,=\,0s$.
}\label{granos}
\end{figure}

\begin{figure}
\centering
\includegraphics[clip,width=\textwidth]{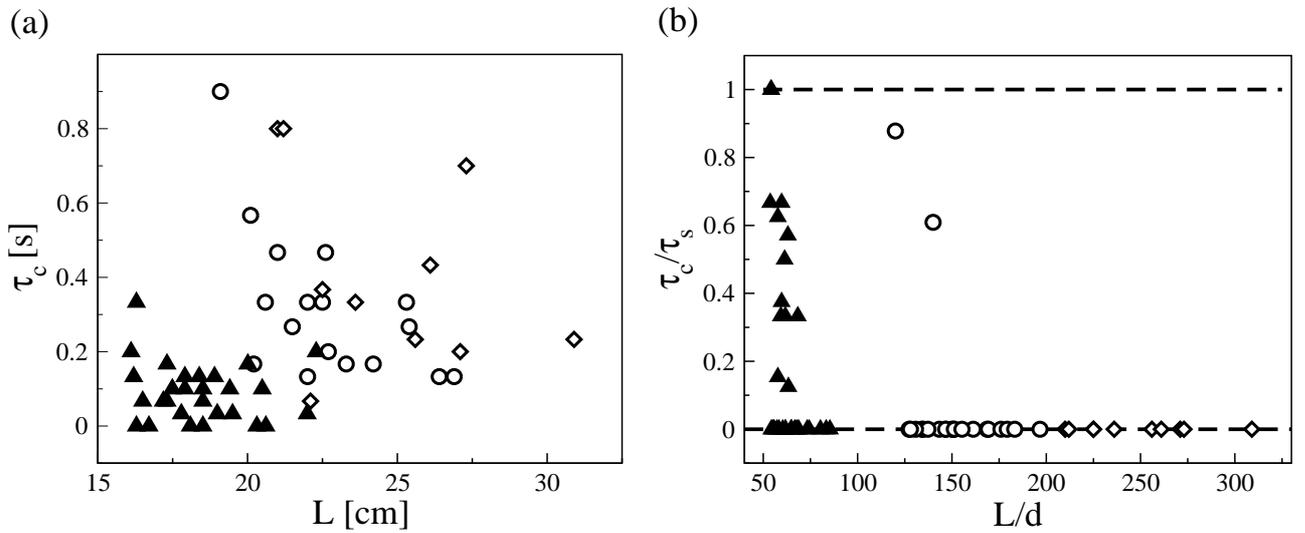}
\caption{Time for the onset of collapse as a function of initial column height. We show (a) $\tau_c$ as a function of the initial height of the granular 
column $L$ and (b) the ratio $\tau_c/\tau_s$ vs. $L/d$, for paper silos with $D\,=\,(4.0\pm0.1)cm$ filled above the critical collapse height, 
with monodisperse glass spheres of different sizes and using different diameters of the discharge orifice:
$d=(1.00\pm 0.05)mm$ with $a\,=\,(0.6\pm0.1)cm$ $(\diamond)$, 
$d=(1.50\pm 0.05)mm$ with $a\,=\,(1.0\pm0.1)cm$ $(\bigcirc)$ and $d=(3.00\pm 0.05)mm$ with $a\,=\,(1.8\pm0.1)cm$ $(\blacktriangle)$. 
In (b), when $\tau_s>\tau_p$ we set $0\equiv\tau_c/\tau_s$. This choice is arbitrary since the large deflection of the silo makes it impossible to follow the
grains and the system is already out of the regime of interest here. However, the choice reflects the fact that in these cases the value of $\tau_s$ is large
compared with the other time scales under study. We find values of $\tau_p$ between $(0.20 \pm 0.03)s$ and $(2.50 \pm 0.03)s$, with higher values 
associated to smaller particles.
The dashed lines show the zero and unity levels. 
Results for different grain sizes form separate groups due to the fact that here we only show events where $L>L_c$ and the critical height depends on $d$ 
\cite{gutierrez-tbp}. 
}\label{tiempos}
\end{figure}

We want to establish when the grains in contact with the silo wall enter a dynamic friction regime, and if the deformation of the wall has an effect on this transition. Therefore we study the movement of the grains in rigid and deformable wall silos. We define the time of onset of sliding, $\tau_s$,
as the time when the grains adjacent to the wall start sliding.\\ 
For PMMA silos, as they are transparent,
we can study the velocity profile for grains adjacent to the wall for the whole column of grains. 
In Fig.\ref{profile} we show the displacement of grains adjacent to the wall in the 
upper layers of the granular column and close to the bottom  of the silo. Within our time resolution,
for the cases of interest here the grains adjacent to the wall start moving simultaneously for all heights above $1.5D$. Below a height approximately
equal to one silo diameter, grains are observed to start moving earlier, flowing inwards and downwards into the bulk as well as sliding against 
the wall up to the stagnant zone, where they form an inverted cone of static material with a slope determined by the angle of repose of the grains.
We have found a similar early flow pattern for grains adjacent to the wall in rigid wall silos and in deformable silos made of
a thin plastic sheet for all grain sizes under study. \\
We are specially interested in granular columns with initial heights close to the collapse height, which is no less than $3D$ for the cases under study. 
Since paper silos are translucent we can only see directly the grains in 
the upper layers of the granular column. The above result allows us to conclude that it is enough to monitor the displacement
of the grains in the upper layers to determine when most of the grains adjacent to the wall start sliding. \\
In Fig.\ref{granos}(a) we show the displacement of the upper layers of grains during the initial instants of the discharge of rigid wall silos
with different apertures and for two different grain sizes. One can resolve an interval of time after the beginning of the discharge
where the grains are not moving relative to the wall. 
The average time of onset of sliding for these rigid wall silos is around $\tau_s=(0.20 \pm 0.03)s$. It is roughly the same for all measurements 
made in rigid silos within our experimental resolution and for the limited range of parameters studied here. 
Not surprisingly, the local average velocities after $\tau_s$ for different values of $a$ and $d$ show a correlation with the corresponding 
value of the flow rate expected from Beverloo's law \cite{beverloo}.\\ 
In Fig.\ref{granos}(b) we observe typical examples of the vertical displacement relative to the wall 
of grains in the upper layers of the granular column during the initial instants of the discharge of deformable paper silos.
In this case, since the wall may deform and move, we follow simultaneously the 
grains and three points painted on the silo wall to compute the relative displacement (Fig.\ref{Exp-Setup}(b)). 
Again, an interval of time where the grains are not moving relative 
to the wall can be resolved.\\
When we start with a column with a height $L$ above the collapse threshold, we observe the behavior of the grains
until the time of collapse,
$\tau_p$. After $\tau_p$ we do not continue monitoring the system as total failure has occurred.
For deformable silos, local average velocities after the grains begin sliding are very variable as well as the time of onset of sliding ($\tau_s$).
In some cases with initial filling heights above $L_c$, it can even occur that the grains begin sliding only after the silo has suffered a
permanent deformation. Typical examples of this behavior can be observed in Fig.\ref{granos}(b), for the largest values of $L-L_c$, 
where no displacement is observed up to $\tau_p$. These cases also show that the time of onset of sliding in deformable wall silos can be much longer
than the average $\tau_s$ measured for rigid wall silos. 
This may be due to the fact that the wall folds in as the diamond shaped localizations develop, these folds may block the flow of grains above the 
deformations (see Fig.(\ref{deformacion})).\\
In Fig.\ref{tiempos}(a) we show the time of onset of collapse, $\tau_c$, for cases when the initial height of the column is larger that $L_c$. 
This quantity shows big fluctuations but
seems to exhibit a tendency to diminish for larger values of $L$. It also seems to depend on the choice of grain and 
orifice size in our experiments. One may speculate that the former responds to the fact that a larger $L$ means that a larger mass of grains is contained in
the silo, and hence it may be easier to produce forces on the wall that are large enough to trigger the collapse. However, more experiments need to be performed 
in order to clarify these observations, which fall beyond the scope of the present work.\\
Finally, we want to establish if the collapse begins before or after the grains start sliding against the wall. In Fig.\ref{tiempos}(b)
we show the ratio of the time of onset of collapse and the time of onset of sliding ($\tau_c/\tau_s$) for various initial filling heights and various 
particle sizes. This ratio can be seen to be always between zero and one, meaning that 
the onset of collapse occurred always before the grains are sliding, i.e, in a static friction regime. Moreover, in most cases the sliding of 
the grains does not occur until far after a permanent deformation of the wall has taken place, specially as the grain size becomes smaller 
and the initial column height becomes higher. \\

\section{Discussion}

The fact that the time of onset of sliding is always longer that the time of onset of collapse indicates that the collapse process is triggered in a regime of
static friction between the grains and the silo wall. As the grains in contact with the wall will eventually enter the flowing region,
the mobilization of friction and hence the force on the wall of the silo
must be evolving during the time before $\tau_s$ towards full mobilization and a maximum global and local force on the wall. 
Moreover, as can be seen in Fig.\ref{tiempos}(b), we have found that in most cases the plastic deformation and complete failure of the structure occur 
before $\tau_s$, and not only the onset of the collapse under the criterion proposed here.\\
Perge et al \cite{perge12} monitored the evolution of the force on the wall and bottom of rigid wall silos before and after the discharge, 
with different conditions of initial friction mobilization at the wall. They showed that Janssen's screening effect \cite{janssen} depends on the mobilization of friction.
Janssen's saturated stress profile may not be found before the discharge when the system is in a state of undetermined
mobilization of friction, but may be produced by fully mobilizing the friction during the preparation of the system or by allowing it to evolve
during the discharge. Previously, Wambaugh et al \cite{wambaugh10} studied the force chains inside a two dimensional silo controlling the initial mobilization
and showed that when mobilization at the wall is not uniform, Janssen's screening effect is not produced. 
From these results regarding the stresses in the silo, even though these where obtained under experimental conditions different from ours, we arrive at the
conclusion that in the cases under study in this paper the collapse is triggered by forces on the wall that may not produce a 
saturating stress profile as would be expected from the classical Janssen model. 
However, by enhancing Janssen's model to take into account fluctuations in static friction mobilization, one may produce a stress profile with a variable screening length,  $\lambda_{scr}$, 
the characteristic depth at which the stress profile is seen to saturate. 
A friction mobilization dependent  $\lambda_{scr}$ could produce a stress profile that is non-saturating at the beginning of the discharge and evolves as mobilization
changes towards the saturating Janssen profile.
Besides, as can be seen in Fig.\ref{deformacion}, the localized nature of the deformations of the silo wall that occur during the discharge, suggests
that friction fluctuations may be playing a crucial role in triggering the deformation. A first step in this direction can be found in
the stochastic version of Janssen's model presented by Berntsen and Ditlevsen \cite{berntsen00}, from which we obtain a mobilization
dependent screening length.
Although a precise connection with our results is lacking, we are convinced that this model will prove useful in understanding discharge-driven collapse of 
silos.\\
In Janssen's model the granular medium is assumed as a continuum and force balance on a horizontal cylindrical slice of granular material reduces to the adimensional equation:
\begin{equation}\label{difEq}
\frac{dP}{dz}+\kappa_{\mu}P=\phi
\end{equation}
where $P=4P_v/\rho g D$ is the normalized vertical pressure and $z=4z'/D$ is the normalized depth, with $\rho$ the density of the grains and $\phi$ 
the packing fraction. The parameter $\kappa_{\mu} \equiv K\mu$ 
involves the Janssen's redirection parameter $K$ of vertical to horizontal stress and an effective
grain-to-wall friction coefficient $\mu$ \cite{janssen}. In the usual Janssen's model the parameter $\kappa_{\mu}$ is assumed 
constant and mobility of friction is assumed to be equal to one everywhere, which leads to an exponentially saturated pressure profile.\\
Berntsen and Ditlevsen \cite{berntsen00} have considered 
a stochastic version of this model by introducing fluctuations in $\kappa_{\mu}$.
The result is the usual Janssen's 
pressure profile, with the only difference being that we must replace $\kappa_{\mu}$ by $m(\Lambda) \langle \kappa_{\mu} \rangle$,
where $\Lambda= \langle \kappa_{\mu} \rangle/\sigma^2$ depends on the average value of $\kappa_{\mu}$ and also on its variance $\sigma^2$.
The function $m(\Lambda)=\Lambda\ln \frac{\Lambda+1}{\Lambda}$ is in the interval $(0,1)$.
The introduction of $\kappa_\mu$ as a random variable can be interpreted as a way to introduce our ignorance about the mobilization
of friction at the wall. In Eq.\ref{difEq} the term $\kappa_{\mu}P$ represents normalized friction per unit area and per unit length. 
From the results of Berntsen and Ditlevsen the following relation can be obtained \cite{berntsen00}:
\begin{equation}\label{mob}
\frac{\langle \kappa_{\mu}P\rangle}{\langle \kappa_{\mu} \rangle \langle P\rangle}= m(\Lambda),
\end{equation}
where  $\langle \kappa_{\mu}P\rangle$ is the expected value of $\kappa_{\mu}P$.
Since $m(\Lambda)\in (0,1)$, we can interpret $m(\Lambda)$ in Eq.\ref{mob} as the average mobilization of friction at the wall. 
Within this picture,
if the fluctuations of $\kappa_{\mu}$ decay then $\Lambda$ increases and, consequently, the average mobilization
of friction increases.
From Eq.\ref{mob} we see that
we can also interpret $m(\Lambda)$ as a measure of correlations between $P$ and $\kappa_{\mu}$, which are related
by Eq.\ref{difEq}. A large average mobilization, which is equivalent to small fluctuations of $\kappa_{\mu}$, implies small correlations between 
$P$ and $\kappa_{\mu}$.
Global fluctuations in the total force on the silo wall are also expected to be important in this context \cite{vanel99,chand12}.\\
From this framework, it is expected that in a regime of static friction the evolution of the average mobilization of friction is equivalent to the
evolution of the average and fluctuations of $\kappa_\mu$ \cite{wambaugh10}. The dimensionless friction mobilization dependent screening length 
$\lambda_{scr}$ obtained from the stochastic model of Berntsen and Ditlevsen becomes $\lambda_{scr}=[m(\Lambda) \langle \kappa_{\mu} \rangle]^{-1}$.
Further work can be done by considering how to couple the evolution of friction mobilization to Eq.\ref{difEq} and
allowing $\phi$ to fluctuate \cite{aste08,aste10}. Correlations between $\kappa_{\mu}$ and $\phi$ \cite{ovarlez03,wambaugh10} may be introduced in such a model, 
and this can be analyzed with the approach of reference \cite{jia00}. This conceptual framework can be useful to link computer simulation results with our
experimental observations. It would be interesting to study how $K$, $\mu$ and $m$ are related.

\section{Conclusion}

We have studied experimentally the time dependent behavior of individual grains in contact with the wall of silos at the beginning of the discharge, 
in rigid and deformable wall silos. In all cases we could resolve an interval of time after the beginning of the discharge where the grains adjacent to
the wall remain static. In all our experiments, the deformation that leads to the collapse of the silo never appears after the grains adjacent to the
wall begin sliding. We conclude that the collapse of the silo is triggered in a regime of static friction between the grains and the wall.

\newpage

\section{Acknowledgements}

We would like to thank C. Mateu for her help with the image processing and I. J. S\'anchez for useful discussions.
We are grateful for the support of PCP-FONACIT for the project titled "Statics and dynamics of granular materials".
This research was supported in part by DID-USB.

\bibliographystyle{unsrt}

\end{document}